\documentclass{aa}
\usepackage{natbib,graphicx}
\newcommand{\excs}{\extracolsep{\fill}} 
\begin{document}
\title{VLT/NACO adaptive optics imaging of the TY CrA system}
\subtitle{A fourth stellar component candidate detected}
\titlerunning{VLT/NACO AO imaging of TY CrA}
\authorrunning{Chauvin et al.}
\author{
        G. Chauvin\inst{1}\and
        A-M. Lagrange\inst{1}\and
	H. Beust\inst{1}\and
        T. Fusco\inst{2}\and
        D. Mouillet\inst{3}\and
	F. Lacombe\inst{4}\and
	E. Gendron\inst{4}\and
	G. Rousset\inst{2}\and
	D. Rouan\inst{4}\and
	W. Brandner\inst{5}\and
	R. Lenzen\inst{5}\and
	N. Hubin\inst{6}\and
	M. Hartung\inst{5}\and
	C. Perrier\inst{1}
}
\offprints{Ga\"el Chauvin \email{gchauvin@obs.ujf-grenoble.fr}}
\institute{
$^{1}$Laboratoire d'Astrophysique, Observatoire de 
Grenoble, 414, Rue de la piscine, Saint-Martin 
d'H\`eres, France\\
                $^{2}$ONERA, BP52, 29 Avenue de la Division Leclerc Ch\^atillon Cedex, France \\
                $^{3}$Laboratoire d'Astrophysique, Observatoire 
Midi-Pyr\'en\'ees, Tarbes, France\\
		$^{4}$Laboratoire D'Etudes Spatiales et d'Instrumentation en Astrophysique, Observatoire de Paris, Bat. 16, 5 Place J. Jansen, 92195 Meudon, France\\ 
		$^{5}$Max-Planck-Institut f\"{u}r Astronomie, K\"onigstuhl 17, D-69117, Heidelberg, Germany\\
		$^{6}$ESO, Karl-Schwarzschild-Str. 2, D-85748, Garching, Germany\\
              }
\date{Received:  / Accepted: }
\abstract{We report the detection of a possible subsolar mass
companion to the triple young system TY\,CrA using the NACO
instrument at the VLT UT4 during its commissioning. Assuming for TY\,CrA a distance similar to that of the close binary system HD\,176386, the
photometric spectral type of this fourth stellar component candidate is consistent with an $\sim$M4 star. We discuss the dynamical stability of this possible quadruple system as well as the possible location of dusty particles inside or outside the system.}
\maketitle
%
%
\section{Introduction}

The CrA dark cloud has been
extensively studied from X ray \citep{neu} to IR \citep{wilk97}
and radio wavelengths as one of the closest star
forming regions. Nearby clouds such as the reflection nebula around R\,CrA greatly help understanding the stellar formation processes and can be used to
compare general tendancies in various star forming regions \citep[see
e.g.][]{chen97}. This small cloud is now known to host dozens
of young stars, mostly low mass stars and brown dwarfs, plus a few
high mass stars. Among the latters, TY\,CrA (B9) is the one of the
brightest at optical and near IR, together with R\,CrA (A5), HD\,176386
(A0) and T\,CrA (F0). As other peculiar massive multiple systems HR\,5999 with a T Tauri companion \citep{stec95} or the eclipsing binary EK\,Cep \citep{clar95b}, TY\,CrA deserves particular attention as it adresses the question of massive star formation.

The distance of the TY CrA system is presently unknown. However, at $\sim57\,\!''$ away, the binary system HD\,176386 shares common proper motion with TY CrA \citep{teix00}. We can then reasonably assume that these two systems are physically bound and then consider for TY CrA the \textit{Hipparcos} distance of $136_{-19}^{+25}$\,pc measured for HD\,176386 (as assumed by \citet{siebe00}).

\begin{table*}[tb]
\caption[]{\centering NACO observations of TY\,CrA and SAO210858: instrumental set-up and observations logs. Camera S27 and S13 respectively provide 27.0 and 
13.4 mas/pixel sampling. DIT and NDIT correspond respectively to the exposure integration time and the number of exposures.}
\centering
\begin{tabular}{llllllll}
\hline\hline\noalign{\smallskip}
Star & Central Wavelength & Bandwidth & DIT & NDIT & Camera & WFS & Dichroic\\
     & ($\mu$m)           & ($\mu$m)  & (s)  &      &      &     &      \\
\noalign{\smallskip}\hline\noalign{\smallskip}
TY\,CrA      & 2.166 & 0.023 & 0.6 & 100 & S27 & VIS & VIS\\
TY\,CrA      & 1.257 & 0.014 & 1   & 25  & S27 & VIS & VIS\\
SAO\,210858 & 2.166 & 0.023 & 0.5 & 50  & S13 & VIS & VIS\\
SAO\,210858 & 1.257 & 0.014 & 1   & 25  & S13 & VIS & VIS\\
\noalign{\smallskip}
\hline
\end{tabular}
\label{setup}
\end{table*}

TY\,CrA has been identified and characterized by \citet{corp94}
as an eclipsing spectroscopic binary (SB) Pre Main Sequence (PMS)
system. Two years later, \citet{casey95} and \citet{corp96} showed that TY\,CrA
is a triple SB system, the first one detected among young
and massive Herbig stars. Eclipsing binary stars are very precious as
they allow a precise determination of the star masses. Coupled to
evolutionary tracks, they help deriving ages for the system. Both mass
and age are crucial parameters for studies of stellar systems dynamics and
stability and provide thus valuable constraints on binary system
early evolution processes \citep[see e.g.][]{prato}. In the case of TY\,CrA, the
 masses have been found to be M$_{1}=3.0\,$M$_\odot$, M$_{2}=1.6\,$M$_\odot$ and M$_{3}=1.2-1.4$\,M$_\odot$ \citep{casey98,corp96}. The secondary and tertiary components are very close
to the primary, with semi major axes of respectively 14\,R$_\odot$
(0.065\,AU) and
1--1.5 AU. Using evolutionary tracks of \citet{swen94,clar95a,danto94}, \citet{casey98} found an age of order 3~Myr for the system, confirming the estimation done by \citet{beu97}. This figure may be compared with the age of 4~Myr recently derived by \citet{palla01}. The TY CrA system exhibits IR excesses indicative of circumstellar (CS) dust \citep{cruz84,wilk85,bibo92}. At least part of this dust is cold, but the CS material has not been resolved sofar.


The richness of the TY\,CrA system encouraged us to conduct Adaptive
Optics (AO) imaging
observations with NACO at the VLT UT4 in the course of the instrument
commissioning. A faint visual additional companion candidate, unknown
sofar, was discovered close to the unresolved triple system. We present in
Sect.~2 the results of the VLT/NACO AO observations. In Sect.~3,
we discuss the photometry of the additional companion candidate to
TY\,CrA to test if this object is bound and to determine the
corresponding stellar parameters. In Sect.~4, we finally discuss the
dynamical status of the possible quadruple system, i.e the possible existence of a \textit{Kozai} resonance between the possible fourth stellar component and the three others. We also discuss the possible location of dusty particles in this system. 

%
%
\section{Observations and results}
Near IR images of the TY\,CrA system (RA=19:01:41, $\delta$=-36:52:34 (J2000),
V=9.4, Spectral Type Herbig Be) have been recorded on March 31$^\mathrm{th}$
2002 with NACO \citep{lag02,len02}. NACO is
composed of NAOS, the first Adaptive Optics System on the
VLT \citep{rou02} and on the other hand of
CONICA, a 1--5 $\mu$m imaging, coronographic, spectroscopic and
polarimetric instrument \citep{len98}.

For these observations, we selected NACO's visible wavefront sensor. With the adequate dichroic, the visible light  was directed to NAOS and all the IR light towards CONICA. The CONICA detector is an Aladdin InSb $1024\times1024$
pixel array. The cameras used were either S13 or S27 which provide
respectively 0.0134 and 0.0270 arcsec/pixel samplings adapted to the observing wavelengths.  
\begin{figure}[!h]
\includegraphics[angle=-90,width=\columnwidth]{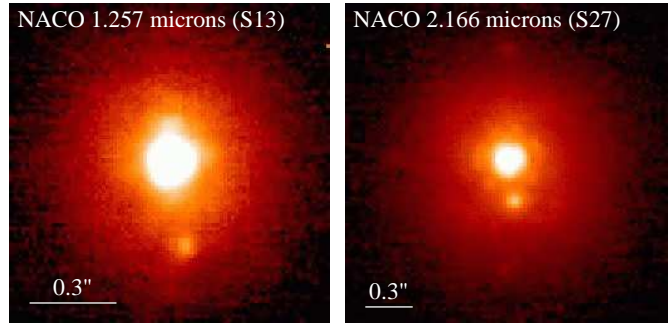}
\caption{VLT/NACO AO imaging of the unresolved triple system TY\,CrA
and the fourth stellar component candidate detected at $0.3\arcsec$ at
$1.257~\mu$m and $2.166~\mu$m. North is up and east is left.}
\label{obs}
\end{figure}
\begin{table}[!h]
\caption[]{MISTRAL deconvolution results. Contrast and relative position
between the unresolved triple system and the fourth stellar component candidate observed at $1.257~\mu$m and $2.166~\mu$m.}
\label{phot}
\begin{tabular*}{\columnwidth}{@{\excs}lllll}
\hline\hline\noalign{\smallskip}
Source & $\Delta\mathrm{[1.257]}$ & $\Delta\mathrm{[2.166]}$ & Separation & P.A.\\
       &  (mag.) & (mag.) & (arcsec) & (degrees)\\
\noalign{\smallskip}\hline\noalign{\smallskip}
TY\,CrA & $3.5\pm0.3$ & $2.8\pm0.1$ & $0.294\pm0.007$ & $188.5\pm1$\\
\noalign{\smallskip}\hline
\end{tabular*}
\end{table}
The star
SAO\,210858 (RA=19:03:12, $\delta$=-37:29:11 (J2000), A5, V=10.0) located close
 to
TY\,CrA was also recorded to get PSF calibrations as close as possible in time to
TY\,CrA observations. The detail of the instrumental set-up is given in
Table~\ref{setup}.  Although the observations were performed under poor
seeing conditions ($1.3$--$1.8\arcsec$), a faint additional object is detected
at $0.3\arcsec$ from the unresolved triple system TY\,CrA. Figure~\ref{obs}
presents the AO images obtained at $1.257~\mu$m and $2.166~\mu$m. The
observations were then deconvolved using MISTRAL algorithm \citep{cona00,fus99}. The results are reported in Table~\ref{phot}.
\section{Companion or background object ?}

As this new companion candidate cannot be identified with one of the known spectroscopic components of the TY\,CrA triple system (contradiction with well known orbit parameters), the first question is whether the photometry of this object is consistent with the photometry of a fourth, gravitationally bound stellar component and if so, to derive the corresponding stellar parameters. The photometry of the unresolved TY\,CrA system
obtained by \citet{deWi01}, $J=7.36\pm0.1$ and $K=6.57\pm0.1$,
and the observed contrasts (see Table~\ref{phot}) are used to determine
the photometry of the new object. We have made the approximation that the contrasts obtained with the narrow band filters at 1.257 and $2.166~\mu$m were respectively similar to the ones with broad bands filters $J$ and $K$. 
\begin{figure}[t]
\includegraphics[width=\columnwidth]{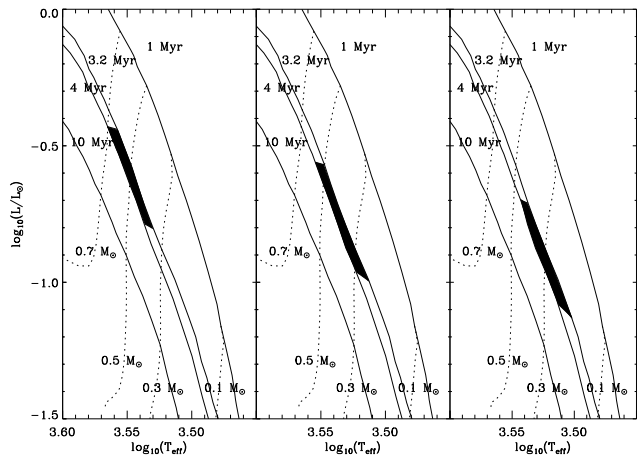}
\caption{Evolutionary tracks of \citet{bar98} with the location of
the fourth stellar component candidate to TY\,CrA based on photometry and assuming an
age between  3.2 and 4~Myr for 3 hypothesis of distance: 161~pc, 136~pc and 117~pc (from left to right).}
\label{fig2}
\end{figure}
\begin{table}[t]
\caption[]{Stellar parameters for the fourth stellar component candidate to TY\,CrA if
bound to the system.}
\label{stell}
\begin{tabular*}{\columnwidth}{@{\excs}llll}
\hline\hline\noalign{\smallskip}
\multicolumn{4}{c}{Age of 3.2~Myr}\\
\noalign{\smallskip}\hline\noalign{\smallskip}
Distance & Distance & Mass & Effective\\
to Earth & to TY\,CrA && temperature\\
(pc) & (AU) & (M$_\odot$) & (K)\\
\noalign{\smallskip}\hline\noalign{\smallskip} 
117 & 35 & [0.2,\,0.4] & [3178,\,3445]  \\
136 & 41 & [0.25,\,0.5]  & [3260,\,3525]   \\
161 & 48 & [0.35,\,0.62] & [3395,\,3621]    \\
\noalign{\smallskip}\hline\hline\noalign{\smallskip}
\multicolumn{4}{c}{Age of 4~Myr}\\
\noalign{\smallskip}\hline\noalign{\smallskip}
Distance & Distance & Mass & Effective\\
to Earth & to TY\,CrA && temperature\\
(pc) & (AU) & (M$_\odot$) & (K)\\
\noalign{\smallskip}\hline\noalign{\smallskip}
117 & 35 & [0.25,\,0.45] & [3270,\,3492]   \\
136 & 41 & [0.3,\,0.57]  & [3340,\,3584]   \\
161 & 48 & [0.4,\,0.7] & [3452,\,3674]    \\
\noalign{\smallskip}\hline
\end{tabular*}
\end{table}
The photometry of the fourth component candidate was then compared to the
evolutionary tracks of \citep{bar98} for an age of 3.2
and 4~Myr. Regarding the important \textit{Hipparcos} parallaxe
uncertainties, three hypotheses of distances were considered: 161~pc, 136~pc
and 117~pc. 

In each case, we used then the color-magnitude diagram for the near-IR color ($J-K$) to determine the expected stellar
parameters for the fourth stellar component candidate to TY\,CrA if gravitationally bound. 
The results are
reported in Table~\ref{stell} and are also overplotted to the
evolutionary tracks in Fig.~\ref{fig2}. For these three distances,
the corresponding spectral types found for an age ranging between
3.2 and 4~Myr are M1--M4, M2--M5 and M3--M6. Further spectroscopic observations
will help to confirm the companionship of this newly detected object and to characterize precisely its spectra. 

\section{Dynamical status of the quadruple system ?}

We cannot state that the newly discovered object is really bound to the triple system even if the measured colour and brightness are consistent in the frame of the distance and age of the system. Repeated astrometric observations are needed to assert this point. One can however address the effects and consequences if it is physically bound. Hereafter, a M$_{4}\sim
0.4\pm0.2\,$M$_\odot$ component orbiting the triple system at $\sim 41$~AU is considered.

A detailed dynamical modeling of \citet{beu97} showed that the stability of the triple system was ensured by tidal
effects within the central binary.  More specifically, if tidal effects
were not present, the secular dynamics of the triple system would be
dominated by the \emph{Kozai} resonance \citep{koz62,har68}. In triple stellar system, this phenomenon is characterized by drastic periodic increases of the eccentricity of the central binary, while in the same time the
mutual inclination $i$ between the orbit of the central binary and that of
the third component would be subject to sharp drops by more than $40\degr$ on very short timescales ($<10^{3}$ yr). Active as soon as the inclination $i$ between the two orbits is $\ga 40\degr$ \citep{sod82}, only the orbit with the smaller angular momentum (i.e the inner orbit) is strongly affected \citep{beu97}. In the case of TY\,CrA, the ratio of the third body angular momentum over the central binary angular momentum is $\sim 4.8$. This is enough for
initiating a very strong \textit{Kozai} resonance which would strongly affect the central binary (with a possible physical collision) in the absence of tidal effects.

If now a fourth member is identified, it is worth wondering whether
a \textit{Kozai} resonance exists between this object and the third body around the central binary. This is of valuable interest, because in that case, no tidal effects in the orbit of the third body are to be expected to assure the dynamical stability of the whole system. Actually a specific dynamical study must be initiated and will be the subject of future work. Depending on the mass assumed for the fourth body, and on the semi-major axis and eccentricities values taken (compatible with the measured projected separation), the angular momentum ratio between the orbits of the third and fourth component is expected to range between $\sim 1$ and
$\sim 4$. If the mutual inclination is high enough, we may expect
a significant \textit{Kozai} resonance as soon as this ratio is $\ga 2$. We thus
cannot definitely conclude at present, but we may suspect that it will be at work and will affect the overall dynamics of the system.

Another important issue is the presence of dust in the system.
\citet{cruz84,wilk85} and \citet{bibo92}
report the detection of a strong far-infrared excess attributed
to cold ($95\pm 10$\,K) dust located at about 900\,AU from the central
binary. According to \citet{bibo92}, the fact that this distance is
very large leads to question whether this dust is of CS origin:
if there is some CS dust, why is it only at large distance ?
If the whole system is quadruple, dust particles are very unlikely to be dynamically stable at close
distance from the central binary. They might be dynamically stable between the central triple system and the fourth body. This actually depends on the orbit assumed for the
fourth component, and on the strength of an eventual \textit{Kozai} resonance.
If present, dust in this area would be warmer ($\ga 200\,$K) and
generate a mid-infrared excess that has not been detected sofar.
In this context, this non-detection is perhaps an indirect indication
of dynamical instability in this region and may put some constraints
on the dynamical status of the system. Dust is more safely expected to be found orbiting the
four-body system, but not closer to 2--3 times the distance of the fourth
body, i.e. at least 100\,AU, and probably more if radiation pressure
from the central binary is taken into account. The structure of this
dust component (circum-quadruple disk or spherical shell ?) is an open
question and should depend actually on the dynamical behavior of the
four-body system. Obviously a dedicated dynamical study is required.

\section{Conclusion}

We have presented results of VLT/NACO adaptive optics imaging of the TY\,CrA system. A fourth stellar component candidate is detected at $\sim0.3\arcsec$ from the unresolved triple system. It is not possible at present time to conclude whether this fourth component is physically bound to the system. If this is the case, this object is likely to be a $\sim$M4 dwarf based on the evolutionary tracks of \citet{bar98} for an age ranging between 3.2 and 4~Myr and a distance of $136_{-19}^{+25}$\,pc. If so, we can then expect the possible existence of a \textit{Kozai} resonance between the orbit of the fourth component and the third body around the central binary, which could affect the stability of the whole system. Dust particles might be also dynamically unstable between the third and fourth component. The only detection of a strong infrared excess attributed to cold dust orbiting the four-body system may be an indirect indication of dynamical instability in this region. Further dynamical study must be initiated to investigate the status of this possible quadruple system and of the possible location of dust inside or outside the system. 

\begin{acknowledgements}
We would like to thank the staff of the ESO paranal observatory and the NAOS/CONICA team for their work on the instrument (without them, this work could not have been carried on) and particulary Fran\c{c}oise Crifo for her precious comments.
\end{acknowledgements}




\begin{thebibliography}{}
%
\bibitem[Baraffe et al.(1998)]{bar98} Baraffe I., Chabrier G., Allard F. \&
Hauschildt P.H., 1998, A\&A 337, 403
%
%
\bibitem[Beust et al.(1997)]{beu97} Beust H., Corporon P., Siess L.,
Forestini M. \& Lagrange A.-M., 1997, A\&A, 320, 478
%
\bibitem[Bibo et al.(1992)]{bibo92} Bibo E.A., Th\'e P.S. \& Dawanas D.N.,
1992, A\&A, 277, 439
%
\bibitem[Casey et al.(1995)]{casey95} Casey B.W., Mathieu, R.D., Suntzeff, N.B. \& Walter, F.M. 1995, AJ, 109, 2156 
%
\bibitem[Casey et al.(1998)]{casey98} Casey B.W., Mathieu, R.D., Luiz Paulo, R.V., Andersen, J. \&  Suntzeff, N.B., 1998, AJ, 115, 1617 
%
\bibitem[Chen et al.(1997)]{chen97} Chen H., Grenfell T.G., Myers P.C. \&
Hughes J.D., 1997, ApJ 478, 295
%
\bibitem[Claret(1995a)]{clar95a} Claret A. 1995, A\&A, 109, 441
%
\bibitem[Claret(1995b)]{clar95b} Claret A., Gim\'enez A. \& Mart\'in E.L. 1995, A\&A, 302, 741
%
\bibitem[Conan et al.(2000)]{cona00} Conan J.-M., Fusco T., Mugnier L.,
et al., 2000, Proc. SPIE, Vol. 4007 (P.L. Wizinowich, Ed.), p.913
%
\bibitem[Corporon et al.(1994)]{corp94} Corporon P., Lagrange A.-M. \&
Bouvier J., 1994, A\&A 282, L21 
%
\bibitem[Corporon et al.(1996)]{corp96} Corporon P., Lagrange A.-M. \&
Beust H.,  1996, A\&A 310, 228
%
\bibitem[Cruz-Gonzalez et al.(1984)]{cruz84} Cruz-Gonzalez I.,
McBreen B. \& Fazio G.G., 1984, ApJ 279, 679
%
\bibitem[D'Antona et al.(1994)]{danto94} D'Antona F. \& Mazzitelli I. 1994, ApJS, 90, 467
%
\bibitem[de Winter et al.(2001)]{deWi01} de Winter D., Van den Ancker M.E., Maira A., et al., 2001, A\&A, 380, 609 
%
\bibitem[Fusco et al.(1999)]{fus99} Fusco T., V\'eran J.-P., Conan J.-M. \&
Mugnier L., 1999, A\&AS 134, 193
%
\bibitem[Harrington(1968)]{har68} Harrington R.S., 1968, AJ 73, 190
%
\bibitem[Kozai(1962)]{koz62} Kozai Y., 1962, AJ 67, 591
%
\bibitem[Lagrange et al(2002)]{lag02} Lagrange A.-M., Mouillet D., Beuzit J.-L., et al., 2002, SPIE, in press 
%
\bibitem[Lenzen et al(1998)]{len98} Lenzen R., Hofmann R., Bizenberger P. \& Tusche A., 1998, SPIE, Vol. 3354
%
\bibitem[Lenzen et al(2002)]{len02} Lenzen R., Hartung M., Brandner et al. 2002, SPIE, Vol. 4841
%
\bibitem[Neuh\"auser et al.(2000)]{neu} Neuh\"auser R., Walter F.M., Covino E., et al. 2000, A\&AS 146, 323
%
%
\bibitem[Palla \&\ Stahler(2001)]{palla01} Palla F., Stahler S.W., 2001
ApJ 553, 299
%
\bibitem[Prato et al.(2002)]{prato} Prato L., Simon M., Mazeh T., et al., 2002, ApJ 569, 863 
%
\bibitem[Rousset et al.(2002)]{rou02} Rousset G., Lacombe F., Puget P. et al., 2002, SPIE, Vol. 4007
%
\bibitem[Siebenmorgen et al.(2000)]{siebe00} Siebenmorgen R., Prusti T., Natte A. \& M\"uller T.G. 2000, A\&A, 361, 258
%
\bibitem[S\"oderhjelm(1982)]{sod82} S\"oderhjelm S., 1982, A\&A 107, 54
%
\bibitem[Stecklum et al.(1995)]{stec95} Stecklum B., Eckart A., Henning Th. \& L\"owe M. 1995, A\&A, 296, 463
%
\bibitem[Swenson et al.(1994)]{swen94} Swenson F.J, Faulkner J., Roger F.J. \& Iglesias C.A. 1994, ApJ 425, 286
%
\bibitem[Teixeira et al.(2000)]{teix00}Teixeira R., Ducourant C., Sartori M.J. et al. 2000, A\&A, 361, 1143
%
\bibitem[Wilking et al.(1985)]{wilk85} Wilking B.A., Harvey P.M.,
Joy M., Hyland A.R. \& Jones T.J. 1985, ApJ 293, 165
%
\bibitem[Wilking et al.(1997)]{wilk97} Wilking B.A., McCaughrean M.J.,
Burton M.G., et al., 1997, AJ 114, 2029
%
\end{thebibliography}
\end{document}